\documentclass{PoS}

\let\ifpdf\relax
\usepackage{amsmath}
\usepackage{amsthm}
\usepackage{amssymb}
\usepackage{amsfonts}
\usepackage{graphicx}
\usepackage{bm}
\usepackage{epsfig}

\usepackage{graphicx}
\usepackage{float}
\usepackage{wrapfig}

\newcommand{\sign}{{\rm sign\,}}
\newcommand{\tr}{{\rm tr\,}}

\newcommand{\SU}{{\rm SU\,}}

\newcommand{\eins}{\leavevmode\hbox{\small1\kern-3.8pt\normalsize1}}
\newcommand{\be}{\begin{eqnarray}}
\newcommand{\ee}{\end{eqnarray}}

\title{Phase Diagram of Wilson and Twisted Mass Fermions at finite isospin chemical potential}

\ShortTitle{Phase Diagram of Wilson and Twisted Mass Fermions at finite isospin chemical potential}

\author{\speaker{M. Kieburg}\\
        Fakult\"at f\"ur Physik, Universit\"at Bielefeld, Postfach 100131, 33501 Bielefeld, Germany\\
        E-mail: \email{mkieburg@physik.uni-bielefeld.de}}

\author{K. Splittorff\\
        Discovery Center, The Niels Bohr Institute, University of Copenhagen, Blegdamsvej 17, DK-2100, Copenhagen \O, Denmark\\
        E-mail: \email{split@nbi.dk}}

\author{J. J. M. Verbaarschot\\
        Department of Physics and Astronomy, State University of New York at Stony Brook, NY 11794-3800, USA\\
        E-mail: \email{jv@chi.physics.sunysb.edu}}

\author{S. Zafeiropoulos\\
        Laboratoire de Physique Corpusculaire, Universit\'e Blaise Pascal, CNRS/IN2P3 
63177 Aubi\`ere Cedex, France\\
 Institut f\"ur Theoretische Physik, Goethe-Universit\"at Frankfurt,\\
Max-von-Laue-Str.~1, 60438 Frankfurt am Main, Germany\\
        E-mail: \email{zafeiro@th.physik.uni-frankfurt.de}}
        
\abstract{Wilson Fermions with untwisted and twisted mass are widely used in lattice simulations. Therefore one important question is whether the twist angle and the lattice spacing affect the phase diagram. We briefly report on the study of the phase diagram of QCD in the parameter space of the degenerate quark masses, isospin chemical potential, lattice spacing, and twist angle by employing chiral perturbation theory. Moreover we calculate the pion masses and their dependence on these four parameters.}

\FullConference{The 32nd International Symposium on Lattice Field Theory\\
		 23-28 June, 2014\\
		 Columbia University New York, NY}

\begin{document}
\section{Introduction}\label{sec:intro}

The simulation of Wilson fermions with twisted quark masses became quite popular in the past decade \cite{latsim}. One reason for choosing a twisted mass is to eliminate zero modes of the massive Dirac operator \cite{twistintr}. Such zero modes are possible for the Wilson-Dirac operator without twist due to the exponential tails of eigenvalues in the gap of the level density \cite{twistintr,DSV10}. Another reason is the order $\mathcal{O}(a)$ improvement \cite{FR04}. However, does the phase diagram change by introducing a twist? In particular, what happens with the phase structures introduced by a finite lattice spacing like the Aoki phase~\cite{Aoki84} and the Sharpe-Singleton first-order scenario~\cite{SS98}?

These questions were analyzed at finite lattice spacing and zero chemical potential \cite{twisted,SW04} and very recently for non-degenerate quark masses \cite{HS14} . The phase diagram changes in both situations. But the changes were never drastic since there were always some relations between the system at no twist and at maximal twist due to the inherent symmetries of the saddlepoint solution of the chiral Lagrangian. Hence one can study the Wilson Dirac operator without a twist by the theory with maximal twist for particular quantities at least in chiral perturbation theory.

In this proceedings we consider the phase diagram as a function of the degenerate quark masses of the up- and down-quark, the isospin chemical potential, the lattice spacing, and the twist angle. The full phase diagram includes pion condensation \cite{SonS01},
the Aoki phase \cite{Aoki84} as well as the Sharpe-Singleton scenario \cite{SS98}.  We briefly summarize the results, details will be given elsewhere \cite{JKSVZ}. In Sec.~\ref{sec:setting} we present the chiral Lagrangian in the $p$-regime. Starting from this point we explain the phase diagram in Sec.~\ref{sec:phases} and the pion masses in Sec.~\ref{sec:masses}. In Sec.~\ref{sec:conc} we summarize and discuss our results.

\section{Chiral Lagrangian in the $p$-regime}\label{sec:setting}

The Wilson-Dirac operator with twisted mass,
\begin{eqnarray}\label{twist-Dirac}
 D_{\omega}=\gamma^\kappa D_\kappa-a \Delta+m\cos\omega+\imath m\sin\omega\gamma_5\tau_3+\mu_{\rm I} \gamma_0\tau_3/2,
\end{eqnarray}
is only $\gamma_5\tau_{1/2}$ Hermitian ($D_{\omega}^\dagger=\gamma_5\tau_{1} D_{\omega}\gamma_5\tau_{1}=\gamma_5\tau_{2}D_{\omega}\gamma_5\tau_{2}$)
if the isospin chemical potential $\mu_{\rm I}$ and the lattice spacing $a$ are chosen real which is the natural case. Here $D_\kappa$ are the covariant derivatives, $\Delta$ is the covariant Laplace operator, $\gamma_\kappa$ are the Euclidean Dirac matrices in the spinor space and $\tau_n$ are the Pauli matrices in the flavor space. The twist angle is chosen to be  $\omega\in[0,\pi/2]$ due to the symmetry $\omega\to\omega+\pi$ and $m\to-m$ as well as the symmetry $\omega\to-\omega$ and $D_\omega\to\gamma_0PD_\omega\gamma_0P$ where P is the parity operator, i.e. $P(D_1,D_2,D_3,D_4)=(-D_1,-D_2,-D_3,D_4)$.

When assuming two physical quarks with identical quark masses $m$ the corresponding effective action in the $p$-regime, in particular in the counting scheme ($V$ is the space-time volume, $p$ is the norm of the momentum, and $\Pi$ the amplitude of the pion fields)
\begin{eqnarray}\label{p-regime}
 m^2,\ \mu^4,\ a^4,\ p^4,\ \Pi^4\propto 1/V,
\end{eqnarray}
is given by \cite{SS98,SW04,HS14,SonS01,chipert}
\begin{eqnarray}
 S_{\rm p}&=&\int_V d^4x\mathcal{L}_{\rm p}(U)=\int_V d^4x\biggl(\frac{F_\pi^2}{4}\tr\widehat{D}_\kappa U\widehat{D}^\kappa U^\dagger-\frac{\Sigma m}{2}\tr(e^{\imath\omega\tau_3}U+e^{-\imath\omega\tau_3}U^\dagger)+a^2C_2\tr^2(U+U^\dagger)\biggl)\nonumber\\ \label{chiLagr-p-regime}
\end{eqnarray}
with $U(x)\in\SU(2)$, $F_\pi$ the pion decay constant, $\Sigma$ the chiral condensate, and $C_2=W_6+W_8/2$ the linear combination of two of the three low energy constants corresponding to the discretization. The covariant derivatives in flavor space are \cite{SonS01,covder}
\begin{eqnarray}
 \widehat{D}_\kappa U=\partial_\kappa U-\mu_{\rm I}[U,\tau_3]\delta_{\kappa0}/2,\quad\widehat{D}^\kappa U^\dagger=\partial_\kappa U^\dagger-\mu_{\rm I}[U^\dagger,\tau_3]\delta_{\kappa0}/2\label{cov-der}
\end{eqnarray}
with $[.,.]$ the commutator. To keep the notation as simple as possible we choose the abbreviations
\begin{eqnarray}
 \widehat{m}=m\Sigma V,\quad \widehat{\mu}_{\rm I}^2=\mu_{\rm I}^2VF_\pi^2,\quad \widehat{a}^2=a^2VC_2,\quad \widehat{p}^2=p^2VF_\pi^2,\quad \widehat{E}^2=E^2VF_\pi^2,\label{dimensionless}
\end{eqnarray}
where $E$ is the energy. Here we underline that the dimensionless lattice spacing $\widehat{a}$ might be imaginary though the physical one, $a$, is always real and positive. The reason is that the low energy constant $C_2$ can be positive or negative. Indeed both  scenarios are possible \cite{SS98,KSV12}. Moreover we consider real as well as imaginary isospin chemical potential to get a better understanding of the phase diagram, particularly of its analytical structure.

In the mean field limit the phase diagram is determined by the modes with zero momentum.
The expansion at the saddlepoint $U_0$ in the pion fields $\widehat{\Pi}(x,t)=1/\sqrt{2\pi}\,\int d^3pdE \,\Pi(p,E)\exp[\imath(p_kx^k-Et)]$ with $\Pi=\Pi^n\tau_n$, i.e. $U=U_0\exp[\imath \widehat{\Pi}/F_\pi]$,  reduces the effective action to
\begin{eqnarray}\label{chiLagexp}
S_{\rm p}&=&V\mathcal{L}_0(U_0)+\int_V d^4x\mathcal{L}_1(U_0,\Pi)+\int_V d^4x\mathcal{L}_2(U_0,\Pi)+o(1).
\end{eqnarray}
The chiral Lagrangian  for the zero momentum modes is given by
\begin{eqnarray}
V \mathcal{L}_0(U_0)&=&\widehat{\mu}_{\rm I}^2\tr [U_0,\tau_3][U_0^\dagger,\tau_3]/16-\widehat{m}\,\tr(e^{\imath\omega\tau_3}U_0+e^{-\imath\omega\tau_3}U_0^\dagger)/2+\widehat{a}^2 \tr^2(U_0+U_0^\dagger).\ \ \ \label{chiLagr}
\end{eqnarray}
At the saddlepoint, the term linear in the pion fields, $\mathcal{L}_1$, vanishes while the term quadratic in the pion fields,
\begin{eqnarray}
 \int_V d^4x\mathcal{L}_2(U_0,\Pi)&=&\frac{1}{F_\pi^2}\sum_{p,E}\left[ \frac{\widehat{p}^2-\widehat{E}^2}{4}\tr\Pi\Pi^\dagger-\frac{\widehat{\mu}_{\rm I} \widehat{E}}{4}\tr[U_0\Pi,\Pi^\dagger U_0^\dagger]\tau_3\right.+\frac{\widehat{\mu}_{\rm I}^2}{16}\tr[\Pi,\tau_3][\Pi^\dagger,U_0^\dagger\tau_3 U_0]\nonumber\\
 &&\hspace*{-3.5cm}\left.+\frac{\widehat{m}}{4}\tr(e^{\imath\omega\tau_3}U_0+U_0^\dagger e^{-\imath\omega\tau_3})\Pi\Pi^\dagger+\widehat{a}^2|\tr(U_0-U_0^\dagger)\Pi|^2-\widehat{a}^2\tr(U_0+U_0^\dagger)\Pi\Pi^\dagger\tr(U_0+U_0^\dagger)\right],\label{L2}
\end{eqnarray}
yields the pion masses. We will return to the pion masses in Sec.~\ref{sec:masses}.

Finally we underline that $\mathcal{L}_0(U_0)-8\widehat{a}^2$ has an exact invariance under $U_0\to\tau_1U_0\tau_2$, $\widehat{a}^2\to-\widehat{a}^2$, $\omega\to\pi/2-\omega$, $\widehat{m}\to-\widehat{m}$, and $\widehat{\mu}_{\rm I}^2\to\widehat{\mu}_{\rm I}^2+32\widehat{a}^2$.  This symmetry will show up in the phase diagram. However the pion masses do not fulfil this invariance since $\mathcal{L}_2(U_0,\Pi)$ does not exhibit this symmetry.

\section{Phase Diagram}\label{sec:phases}

\begin{table}[tbp] \centering
\begin{tabular}[c]{c||c||c}
 phase & region & order parameters \\ 
 \noalign{\vskip\doublerulesep\hrule height 2pt} $I$, & 
 $\widehat{\mu}_{\rm I}^2\geq2|\widehat{m}|\sin\omega\geq0$ (if $\omega=0$ then $\widehat{\mu}_{\rm I}^2>0$) & $\Sigma',C_{\pi_0}\propto\widehat{m}$, \\ 
  $\omega\in[0,\pi/2]$ & and   $\displaystyle \overset{\ }{ 16\widehat{a}^2+\frac{\widehat{\mu}_{\rm I}^2}{2}}\geq\overset{\ }{\frac{\widehat{\mu}_{\rm I}^2|\widehat{m}|\cos\omega}{\sqrt{\widehat{\mu}_{\rm I}^4-4\widehat{m}^2\sin^2\omega}}}\geq0$ & $n_{\rm I}\propto\widehat{\mu}_{\rm I}$ \\ \hline
 $II_{\pm}$, & $\widehat{\mu}_{\rm I}^2\geq2|\widehat{m}|\sin\omega\geq0$ and $\displaystyle\overset{\ }{\frac{|\widehat{m}||\widehat{\mu}_{\rm I}|^2\cos\omega}{\sqrt{\widehat{\mu}_{\rm I}^4-4\widehat{m}^2\sin^2\omega}}> 16\widehat{a}^2+\frac{\widehat{\mu}_{\rm I}^2}{2}}$; & $\Sigma',C_{\pi_0}\propto{\rm sign}\,\widehat{m}$, \\  
 $\omega\in]0,\pi/2[$ & {\bf or} $2|\widehat{m}|\sin\omega> \widehat{\mu}_{\rm I}^2$ & $n_{\rm I}=0$ \\ \noalign{\vskip\doublerulesep\hrule height 2pt}
 $III_\pm^{\omega=0}$ &  $\widehat{\mu}_{\rm I}^2<0\ {\rm and}\ 16\widehat{a}^2>|\widehat{m}|>0$ & $\Sigma'\propto\widehat{m},\ n_{\rm I}=0,$ \\ 
 & & $C_{\pi_0}\propto{\rm sign}\,\widehat{m}$ \\ \hline
  $IV_\pm^{\omega=0}$ & $\widehat{\mu}_{\rm I}^2\geq0\ {\rm and}\ |\widehat{m}|> 16\widehat{a}^2+\widehat{\mu}_{\rm I}^2/2$; & $C_{\pi_0},\ n_{\rm I}=0,$ \\ 
  & {\bf or} $0> \widehat{\mu}_{\rm I}^2\ {\rm and}\ |\widehat{m}|>16\widehat{a}^2$ & $\Sigma'\propto{\rm sign}\,\widehat{m}$  \\  \hline 
 Aoki phase & $\widehat{\mu}_{\rm I}^2=0\ {\rm and}\ 16\widehat{a}^2>|\widehat{m}|>0$ & $C_{\pi_0},n_{\rm I}=0,$\\
 & & $\Sigma'\propto\widehat{m},\ \Delta C_{\pi_0}\neq0$  \\ \noalign{\vskip\doublerulesep\hrule height 2pt} 
  $III_\pm^{\omega=\pi/2}$ & $\widehat{\mu}_{\rm I}^2\leq0\ {\rm and}\ -16\widehat{a}^2>|\widehat{m}|>0$ & $C_{\pi_0}\propto\widehat{m},\ n_{\rm I}=0,$ \\ 
  & {\bf or} $\displaystyle -16\widehat{a}^2>|\widehat{m}|>0\ {\rm and}\ -32\widehat{a}^2>\widehat{\mu}_{\rm I}^2>0$ & $\Sigma'\propto{\rm sign}\,\widehat{m}$ \\ \hline
   $IV_\pm^{\omega=\pi/2}$ & $\widehat{\mu}_{\rm I}^2\leq0\ {\rm and}\ |\widehat{m}|>-16\widehat{a}^2$; & $\Sigma',\ n_{\rm I}=0,$ \\ 
   & {\bf or} $2|\widehat{m}|>\widehat{\mu}_{\rm I}^2>0\ {\rm and}\ |\widehat{m}|>-16\widehat{a}^2$ & $C_{\pi_0}\propto{\rm sign}\,\widehat{m}$
\end{tabular}
\caption{ Regions and order parameters of the phases for small $\widehat{m}$, $\widehat{\mu}_{\rm I}$, $\widehat{a}$. The phases $II_\pm$ split into the regions $III_\pm$ and $IV_\pm$ at the boundaries $\omega=0,\pi/2$. The variance of the $\pi_0$ condensate $\Delta C_{\pi_0}$ does not vanish in the Aoki phase indicating the spontaneous breaking of flavor symmetry and parity.}\label{t1}
\end{table}

To derive the phase diagram we choose the parametrization $U_0=\alpha\eins_2+\imath \beta_n\tau_n$ ($\alpha^2+\beta^2=1$). Minimizing the Lagrangian in these coordinates we find the phases $I$, $III_\pm^{\omega=0}$ and $IV_\pm^{\omega=0}$ for $\omega=0$ and  the phases $I$, $III_\pm^{\omega=\pi/2}$ and $IV_\pm^{\omega=\pi/2}$ for maximal twist $\omega=\pi/2$ while for finite twist $0<\omega<\pi/2$  there are only the phases $I$ and $II_\pm$. The regions of these phases and the behavior of the order parameters (chiral condensate $\Sigma'=\langle\bar{\psi}\psi\rangle$, $\pi_0$ condensate $C_{\pi_0}=\langle\bar{\psi}\gamma_5\tau_3\psi\rangle$, isospin charge density $n_{\rm I}=\langle\bar{\psi}\gamma_0\tau_3\psi\rangle$) are summarized in table~\ref{t1}. The sources of the order parameters are the quark mass, the twisted mass, and the isospin chemical potential, respectively. The condensate of the charged pions $C_{\pi_\pm}=\langle\bar{\psi}\gamma_5(\tau_1\pm\imath\tau_2)\psi\rangle$ plays only a role for the phase $I$ and the Aoki phase but completely vanishes for the other phases. In the phase $I$ and the Aoki phase we have a spontaneous breaking resulting in a massless Goldstone boson which was also found in Refs.~\cite{SonS01,KS02}.

Let us start with the Aoki phase. This phase only exists at $\widehat{\mu}_{\rm I}=\omega=0$. At finite isospin chemical potential and finite twist it appears as a first order phase transition from $I$ to $III_\pm^{\omega=0}$ when varying $\widehat{\mu}_{\rm I}$ and $\omega$ while one enters the Aoki phase through second order end points in the $\widehat{\mu}_{\rm I}=\omega=0$ plane when varying the quark mass or the lattice spacing. A similar statement can be also found in Ref.~\cite{SW04} where the Aoki phase showed up on a lower dimenional cut for $\omega=0$. In our case  the $\pi_0$ condensate jumps from zero ($\mu$ real) to $C_{\pi_0}=2\Sigma\sign\widehat{m}\sqrt{1-\left(\widehat{m}/16\widehat{a}^2\right)^2}$ ($\mu$ imaginary) at the Aoki phase.  Interestingly the Aoki phase has an analogue at imaginary effective lattice spacing ($\widehat{a}^2<0$) and maximal twist ($\omega=\pi/2$) with real isospin chemical potential $\widehat{\mu}_{\rm I}^2=-32\widehat{a}^2$ which exhibits similar behaviors only that now the chiral condensate instead of the $\pi_0$ condensate is discontinuous.

The phase $I$ is the only phase which exists at all twist angles $\omega\in[0,\pi/2]$. It is characterized by a linear dependence of the chiral condensate and the $\pi_0$ condensate on the quark mass, i.e. $\Sigma'\propto\widehat{m}\cos\omega$ and $C_{\pi_0}\propto\widehat{m}\sin\omega$, and a non-linear dependence of the isospin charge density on the isospin chemical potential. Both kinds of dependencies are completely different for the phases $II_\pm$ which only exists for the angles $0<\omega<\pi/2$. In these phases we have no isospin charge density and the chiral condensate and the $\pi_0$ condensate are proportional to the sign of the quark mass but have otherwise a non-trivial dependency on $|\widehat{m}|$. In this way the jump of the chiral condensate when  the quark mass crosses the origin (the  Sharpe-Singleton scenario \cite{SS98}) carries over to finite twist.

The phases $II_\pm$ split at vanishing twist $\omega=0$ and maximal twist $\omega=\pi/2$ to the phases $III_\pm^{0=,\pi/2}$ and $IV_\pm^{\omega=0,\pi/2}$. In all four phases the isospin charge density vanishes as it was the case for $II_\pm$. However the jumps of the chiral and $\pi_0$ condensates  are now different. Only one of the condensates jumps while the other has a linear dependence on the quark mass. A jump of the chiral condensate can be found for the phases $IV_\pm^{\omega=0}$ and $III_\pm^{\omega=\pi/2}$ while the $\pi_0$ condensate jumps for $III_\pm^{\omega=0}$ and $IV_\pm^{\omega=\pi/2}$.

A detailed discussion of the phase diagram, plots and explicit expressions for the condensates, especially the chiral, $\pi_0$, and $\pi_\pm$ condensates and the isospin charge density, will be given elsewhere \cite{JKSVZ}.

\section{Pion Masses}\label{sec:masses}

For the pion masses we have to insert the saddlepoint  of the corresponding phases in the action~\eqref{L2} and solve the equation
\begin{equation}\label{eq1}
\det H(\widehat{E}^2)=0.
\end{equation}
Here the matrix in the determinant is
\begin{eqnarray}
 H(\widehat{E}^2)&=&\left(-\widehat{E}^2+\widehat{\mu}_{\rm I}^2(2\beta_\bot^2-1)+2\widehat{m}(\alpha\cos\omega-\beta_3\sin\omega)-32\widehat{a}^2\alpha^2\right)\eins_3\label{matrix}\\
&&\hspace*{-1cm}+\left[\begin{array}{ccc}  32\widehat{a}^2\beta_\bot^2 & 2\imath\widehat{\mu}_{\rm I}\widehat{E} (1-\beta_\bot^2) & ((\widehat{\mu}_{\rm I}^2+32\widehat{a}^2)\beta_3 +2\imath\alpha\widehat{\mu}_{\rm I}\widehat{E})\beta_\bot \\ -2\imath\widehat{\mu}_{\rm I}\widehat{E}  (1-\beta_\bot^2)  & 0 & (-\widehat{\mu}_{\rm I}^2\alpha+2\imath\widehat{\mu}\widehat{E}\beta_3)\beta_\bot \\  ((\widehat{\mu}_{\rm I}^2+32\widehat{a}^2)\beta_3 -2\imath\alpha\widehat{\mu}_{\rm I}\widehat{E})\beta_\bot & (-\widehat{\mu}_{\rm I}^2\alpha-2\imath\widehat{\mu}_{\rm I}\widehat{E}\beta_3)\beta_\bot & \widehat{\mu}_{\rm I}^2(1-2\beta_\bot^2)+32\widehat{a}^2\beta_3^2 \end{array}\right]\nonumber
\end{eqnarray}
with $\beta_\bot^2=\beta_1^2+\beta_2^2$.

In the phase $I$ we have
\begin{equation}\label{values_I}
 \alpha=\frac{2\widehat{m}\cos\omega}{32\widehat{a}^2+\widehat{\mu}_{\rm I}^2},\ \beta_3=-\frac{2\widehat{m}\sin\omega}{\widehat{\mu}_{\rm I}^2},\ \beta_\bot=\sqrt{1-\alpha^2-\beta_3^2},
\end{equation}
from which one can show that one massless Goldstone mode $\widehat{m}_{\pi,1}$ always exists. The orientation of this mode reads in terms of Pauli matrices
\begin{eqnarray}\label{mode1}
\Pi_{\pi,1}\propto\frac{2\widehat{m}\sin\omega}{\widehat{\mu}_{\rm I}^2}\tau_1+\frac{2\widehat{m}\cos\omega}{32\widehat{a}^2+\widehat{\mu}_{\rm I}^2}\tau_2+\sqrt{1-\frac{4\widehat{m}^2\cos^2\omega}{(32\widehat{a}^2+\widehat{\mu}_{\rm I}^2)^2}-\frac{4\widehat{m}^2\sin^2\omega}{\widehat{\mu}_{\rm I}^4}}\tau_3,
\end{eqnarray}
and becomes a linear combination of the two charged pions $\pi_+$ and $\pi_-$ on most of the phase boundaries. The other two modes have non-degenerate but rather complicated pion masses \cite{JKSVZ}.

\begin{figure}[!t]
 \centerline{\includegraphics[width=1\textwidth]{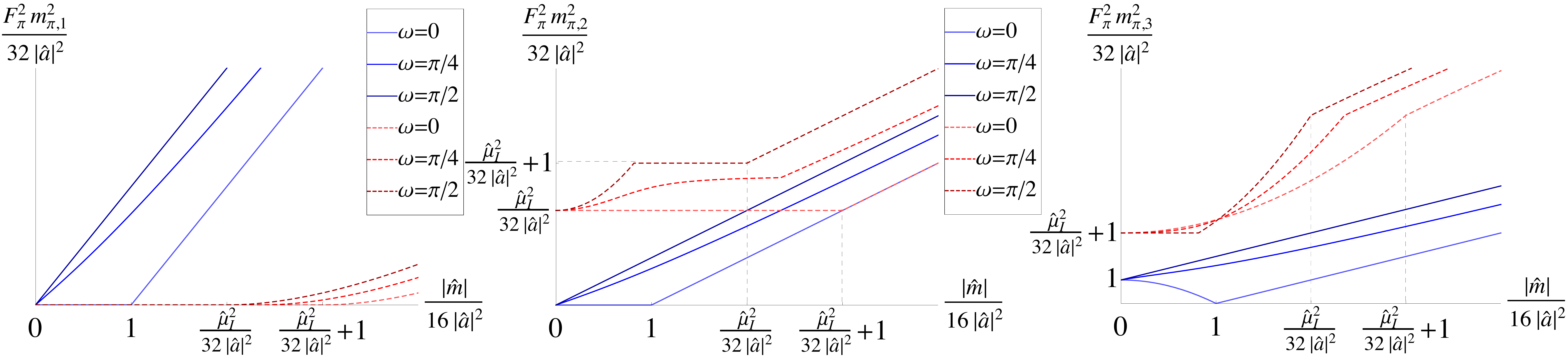}}
\caption{Pion masses at real effective lattice spacing ($\widehat{a}^2>0$)  as a function of the quark mass for various twist angles $\omega$.  Shown is the generic behaviour at vanishing isospin chemical potential (blue, solid curves) and at real isospin chemical potential (red, dashed curves). The vertical grey, dashed lines indicate the positions of the phase transitions between the phases $I$ and $IV_\pm$ at zero and maximal twist. The transitions between the phases $I$ and $II_\pm$ at finite twist angle is always between those two points. The kinks of the masses below these positions result from exact crossings of the masses.}
\label{fig1}
\end{figure}

For the other six phases one can show that the saddle point solution for the variable $\alpha$ is
\begin{equation}\label{values_II}
 \alpha=\sign\widehat{m}\int_0^1\Theta\left(|\widehat{m}|(\sqrt{1-y^2}\cos\omega-y\sin\omega)-16\widehat{a}^2y\sqrt{1-y^2}\right)dy,
\end{equation}
$\beta_3=-\sign\widehat{m}\sqrt{1-\alpha^2}$, and  $\beta_\bot=0$, where $\Theta$ is the Heaviside step-function. Then we find the standard pions as the eigenmodes of the matrix~\eqref{matrix} with the dimensionless masses
\begin{eqnarray}\label{masses-II}
\widehat{m}_{\pi_0}&=&\sqrt{2|\widehat{m}|(|\alpha|\cos\omega+\sqrt{1-\alpha^2}\sin\omega)+32\widehat{a}^2(1-2\alpha^2)},\\
\widehat{m}_{\pi_\pm}&=&\sqrt{2|\widehat{m}|(|\alpha|\cos\omega+\sqrt{1-\alpha^2}\sin\omega)-32\widehat{a}^2\alpha^2}\,\pm\,\widehat{\mu}_{\rm I}.\nonumber
\end{eqnarray}

We plotted the mass dependence at real effective lattice spacing ($\widehat{a}^2>0$) or equivalently $C_2>0$ for all three masses in Fig.~\ref{fig1}. Thereby we considered vanishing and real isospin chemical potential and different values of the twist angle to illustrate the generic behavior. We emphasize that some kinks do not result from phase transitions but from the ordering of the masses $m_{\pi,1}\leq m_{\pi,2}\leq m_{\pi,3}$ and exact crossings of the pion masses.

\section{Conclusions and Outlook}\label{sec:conc}

We presented a brief summary of the phase diagram of two-flavor QCD with a twisted mass, a finite lattice spacing, and a finite isospin chemical potential. Here we considered real as well as imaginary isospin chemical potential to understand the analytical properties of the Dirac operator. A detailed derivation will be  given elsewhere \cite{JKSVZ}.

The phase diagram exhibits the phase denoted by $I$, $II_\pm$, $III_\pm^{\omega=0,\pi/2}$, and $IV_\pm^{\omega=0,\pi/2}$. Only the phases $I$ and $II_\pm$ are present for a finite twist angle $0<\omega<\pi/2$. The other phases only exist in the case of no twist ($\omega=0$) and maximal twist ($\omega=\pi/2$). The phase diagram was derived for the leading order chiral Lagrangian. 
Thus we may expect corrections when some of the parameters like the quark mass or the isospin chemical potential become too large. Limiting cases of these results agree with previous works~\cite{SS98,SonS01,KS02,twisted,SW04}.

Since the chiral Lagrangian of the pion modes with zero momentum exhibits an exact symmetry relating the theory at no twist with the one at maximal twist, the phase diagram and the condensates ($\pi_0$, $\pi_\pm$, and chiral condensate and the isospin charge density) can be studied at maximal twist without any problems and traced back to the original theory. This statement does not hold for the pion masses because the lowest order of the chiral Lagrangian for the pions with non-zero momentum does not reflect this symmetry. In one of the phases, we have one massless Goldstone boson which was also found in Refs.~\cite{SonS01,KS02}. This mode and the other modes are a mix of the standard pion modes. We can identify the $\pi_0$ mode and the charged pions $\pi_\pm$ only in the other phases. 

Surprisingly, the Aoki phase appears as first order phase transition at finite isospin chemical potential and twist. Only in the $\widehat{\mu}_{\rm I}=\omega=0$ plane we find the second order phase transition into the Aoki phase which generalizes the observation of Ref.~\cite{SW04}. At the Aoki phase the $\pi_0$ condensate jumps when the isospin chemical potential switches from a real value to an imaginary one. This behaviour shows that one has to be really careful by interpreting the analytical structure of the phase diagram.

\section*{Acknowledgement}

MK and SZ acknowledge financial support by the Alexander-von-Humboldt Foundation, KS  by the Sapere Aude program of the Danish Council for Independent Research, and JV by U.S. DOE Grant No. DE-FG- 88ER40388. We thank D. P. Horkel and S. Sharpe for fruitful discussions and sharing with us a preprint of Ref.~\cite{HS14}.

\end{document}